# Electronic, Elastic and X-ray spectroscopic properties of direct and inverse full Heusler compounds: DFT+U study

Lalrinkima,[1] Gennady M. Mikhailov,[2] and D. P. Rai[1, *]

[1]*Department of Physics, Pachhunga University Collge, Aizawl-796001, India*
[2]*Institute of Microelectronics Technology and High Purity Materials RAS, 142432 Chernogolovka, Russia*

Half-metallicity, low magnetic damping and high curie temperature ($T_C$) are crucial for application in spintronics and full Heusler alloys in this regard exhibit remarkable properties. Herein, we have considered $Co_2FeAl$ (CFA) and $Fe_2CoAl$ (FCA) as a representative of direct and indirect full Heusler compounds which crystallizes in $L2_1$ and $C1_b$ phases, respectively. The theory of $L2_1$ type Heusler alloys has been well established, however the fundamental understanding of $Fe_2CoAl$ is still under developed. In this paper, we have employed density functional theory (DFT) to study the electronic, elastic and X-ray spectroscopic properties of $Co_2FeAl$ and $Fe_2CoAl$. The electron exchange correlation were treated within a generalized gradient aproximation (GGA) as PBE-scheme. Inorder to include the impact of valence electrons an onsite Coulomb potential is added to GGA as GGA+U. Within both GGA and GGA+U, CFA shows a half-metallic behaviour but FCA is metallic. The calculated values of magnetic moment and $T_C$ are in close agreement with the experimental data

PACS numbers: Valid PACS appear here???

## I. INTRODUCTION

The recent challenges for the material scientists and material engineers is to develop a new and eco-friendly material for the advancement of smart devices. Material science is an open field to accept the new approaches for the technological development. The conventional semiconductor electronic devices works on the basis of electron charge transfer that accomplished by high level of energy consumption. Alternatively, the usage of electron spin-based innovation called spintronics[1–6] which utilizes the spin degrees of freedom, can give extra functionality and new capabilities, including faster switching time and lower power consumption. The spintronic devices solely rely on the spin polarization at the Fermi energy ($E_F$) of the material[7–14]. Researchers believed that the future spin based electronic devices outclass the customary charge-based electronic devices in terms of low power consumption, ecologically and efficiency. The spin functionality of electron was successfully implemented in a device as giant-magneto-resistance (GMR) in 1988[2]. One of the potential competitors is transition based Heusler compounds due to the presence of $d$-orbitals which play an important role in determining the multi-functional properties and their application in diverse field[7]. The transition metal based Heusler alloys show high magnetic moment in the absence of applied magnetic field[15–20]. The other fascinating feature is their high Curie temperature $T_C$ which is an important criteria for technological application and can be well fabricated in devices preserving the ferromagnetism above the room temperature (RT). The theoretical $T_C$ of some of the well known Co-based Heusler compounds calculated from atomistic spherical wave approximation (ASW)[21] are $Co_2VGa(T_C=368$ K), $Co_2CrGa(T_C=362$ K), $Co_2MnAl(T_C=609$ K), $Co_2MnSi(T_C=990$ K), $Co_2MnSn(T_C=899$ K) and $Co_2FeSi(T_C=1183$ K). While the experimental[22–25] values of $T_C$ for the aforementioned compounds are 352 K, 495 K[26], 697 K, 985 K, 1100 K, respectively. Majority of Heusler compounds show high spin polarization at Fermi level ($E_F$), due to the presence of conducting electrons that gives dispersed bands around the $E_F$ at one of the spin channels whereas other spin channel is semiconducting with a band gap. The Half-metallic characteristics in Heusler compound is attributed to the hybridization between the $d - d$ orbitals of the transition elements[27]. The half-metallic Heusler compounds are superior interms of their technological application as compared to other class of half metals like oxides ($CrO_2$, $Fe_3O_4$)[28], manganites $La_{0.7}Sr_{0.3}MnO_3$[28], perovskite $Sr_2FeReO_6$[29,30], pyrites $CoS_2$[31], chalcogenides $CrSe$[32], pnictides $CrAs$[32] etc., due to higher value of $T_C$ and high magnetic moment. The above discussed diverse properties of Heusler compound are highly plausible for the advancement of new spin-based technology[33,34]. However, the understanding of complex magnetization due to the presence of d-electrons and the distribution of magnetic moments in atomistic-scale is crucial. An insight of the intrinsic magnetic properties can be achieved by using magneto-optical spectroscopy, e.g, the X-ray absorption spectroscopy (XAS) and X-ray magnetic circular dichroism (XMCD)[35–38]. The magneto-optical analysis provide a quantitative information about the local spin and orbital magnetic moments of the 3d atoms in the form of $2p_{1/2,3/2} \to 3d$ transitions as $L_{2,3}$ absorption edges. The transition metal based Heusler compound constituted strongly correlated electrons which are not accounted within a conventional density functional theory (DFT). Hence, the information obtained are inaccurate and inadequate. In this paper we have implemented DFT+U (Coulomb potential) approach to integrate the impact of the valence electrons(strongly correlated d-electrons) to procure precision in the outcomes.

## II. COMPUTATIONAL DETAIL

In general, the full Heusler alloy (FHA) is expressed as $X_2YZ$, where X and Y are transition elements and Z species is $p$-elements. There are two types of full-Heusler;direct and inverse depending on the arrangement of atoms according to their electronegativity. The direct FHA crystallizes in $L2_1$ structure, consisting of four inter-penetrating FCC lattices with space group $225(Fm3m)$[39]. Inverse FHA crystallizes in $C1_b$ type structure having space group $216(F\text{-}43m)$ where the electronegativities of X and Y are just the reverse as compared to direct one [Fig.1(a,b)]. In this work, we have used an efficient and globally adopted an open source DFT-package called Quantum Espresso[40]. The relaxed structure at its ground state has been achieved by optimizing the lattice parameters based on Murnaghan's equation of states (EOS)[41]. For further confirmation of thermodynamical stability we have also calculated the frequency dependent phonon dispersion based on linear response method[42] in combination with the density functional perturbation theory (DFPT) as program in Quantum Espresso[43]. Quantum Espresso works on the basis of projected augmented wave (PAW) method that rely on the ultrasoft pseudopotential functional[44]. All electrons are treated by using a generalised gradient approximation (GGA) within Perdew-Burke-Ernzerhof (PBE) functional[45]. A cut-off energy of 50 eV has been considered within a plane wave basis set. The k-points correspond to the electronic wave functions are integrated within a first Brillouin Zone (BZ) by Monkhorst-pack $8\times8\times8$[46]. For the accurate and efficient treatment of the electron-electron interaction among the strongly correlated d-electrons, we have also deployed Hubbard on-site Coulomb interaction (U) within the DFT formalism as GGA+U. The respective U values[47] of Co and Fe are $U_{Co}$= 3.89 eV and $U_{Fe}$= 3.82 eV, have been adopted for our calculation. We have also computed the X-ray Absorption Spectra (XAS) and X-ray Magnetic Circular Dichroism (XMCD) using the spin-polarized, relativistic Korringa-Kohn-Rostoker method (SPR-KKR)[48,49].

## III. RESULTS AND DISCUSSION

The optimized lattice constants ($a_o$) of $Co_2FeAl$ calculated from from GGA and GGA+U are 5.044Å and 5.7036Å respectively. Also, the respective values of lattice constants for $Fe_2CoAl$ within GGA and GGA+U are 5.70Å and 5.73Å. The calculated lattice parameters in comparison with the previously reported theoretical and experimental data are presented in TableI. Our calculated lattice parameters are in qualitative and in close agreement with the previous results.

| Compounds | B*(GPa) | B**(GPa) | a (Å) | $a_{Theo}^{Rep}$(Å) | $a_{Expt}$(Å) |
|---|---|---|---|---|---|
| CFA-GGA | 193.7 | 209.27 | 5.704 | 5.69[39], 5.70[53] | 5.74[51] |
| CFA-GGA+U | 110.4 | | 5.746 | | |
| FCA-GGA | 187.2 | 208.67 | 5.703 | 5.71[39],5.67[54] | 5.766[52] |
| FCA-GGA+U | 325.0 | | 5.733 | | |

TABLE I. Calculated Lattice Constant $a$ (Å) in comparison with the reported data and the Bulk Modulus (where B* based on Murnaghan's equation[41] and B** from Elastic code[50])

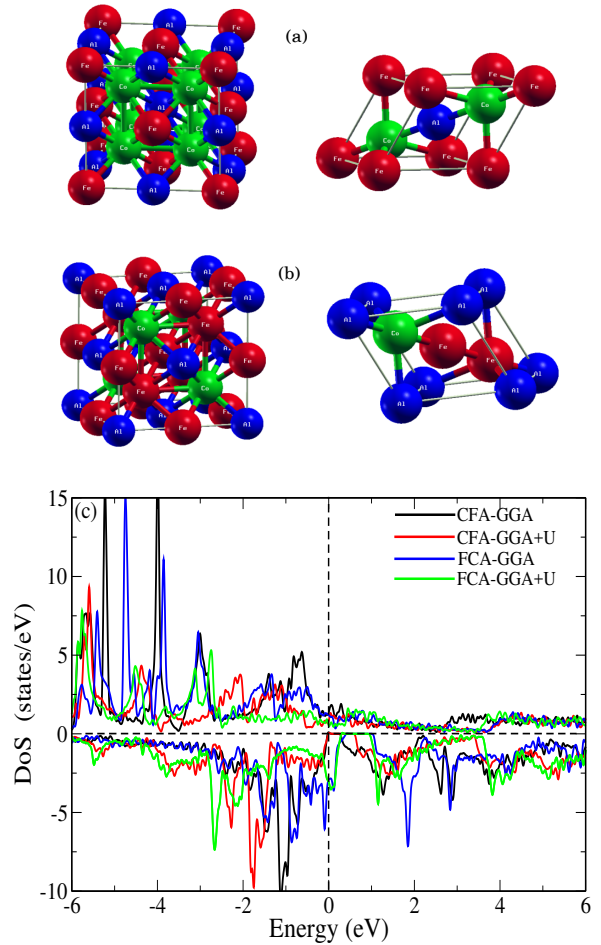

FIG. 1. Conventional and primitive cell of (a) $L2_1$-type $Co_2FeAl$, (b) $C1_b$-type $Fe_2CoAl$ structure and (c) Total density of states (DOS) of $Co_2FeAl$ and $Fe_2CoAl$ calculated from GGA and GGA+U

### A. Electronic and Magnetic properties

By using the optimized lattice constant, the electronic and magnetic properties were calculated. The vivid description of the electronic properties of crystalline material is directly related to density of states(DoS) and band structures. Hence, the total DoS of $Co_2FeAl$ and $Fe_2CoAl$ calculated from GGA and GGA+U are shown in Fig.1(c). We have found that the total DoS of $Co_2FeAl$ and $Fe_2CoAl$ are not comparable despite of having similar chemical composition, due to the differences in their crystallization (space group). In case of $Fe_2CoAl$ the Fermi energy ($E_F$) is completely shifted to the valence region, whereas in $Co_2FeAl$ we can see $E_F$ falling exactly in the band gap in the spin down channel within both GGA and GGA+U approximation. The former exhibits a pure metallic behaviour on the other hand the later gives the half-metallic characteristics. The half metallicity is basically due to the presence of metallic state at one of the spin channels whereas other channel is semiconducting[55]. In $Co_2FeAl$, Co-$d(\uparrow)$ ($d$-$e_g$+$d$- $t_{2g}$) states are dispersed around the $E_F$ and Fe-d($\uparrow$) states are located at around -4.0 eV to -6.0 eV in spin up channel, and a presence of clear band gap at the spin down channel gives all the credibility to its half-metallic behaviour, [Fig.2(a, b)]. The formation of band gap in the spin down channel of $Co_2FeAl$ ascribed to the hybridization between the $d$-orbitals of transtion metal atoms [Fig. 2(a,b)]. The $d-d$ hybridization in $L2_1$ type structure has already been discussed somewhere else[27]. The $d-d$ hybridization gives bonding (occupied) and anti-bonding (unoccupied) states where bonding states are stable with low energy and shifted below the $E_F$. The energy difference between the boniding and anti-bonding states is a measure of half-metallic band gap ($E_g$). The calculated band gap of $Co_2FeAl$ along K-point from GGA and GGA+U are 0.2 eV and 0.51 eV respectively [see Fig. 3(a,b)]. However, the GGA band gap of $Co_2FeAl$ is slightly narrower than that of GGA+U. The degenerated band gap is due to fact that GGA is insufficient in deriving the free electrons efficiently. Rather, the free electrons in the interstitial region are treated as impurities which induced extra states at the edge of the band gap. As a consequence of metal-semiconductor hybrids the $E_F$ lies within the band gap, thus $Co_2FeAl$ exhibit 100% spin polarization. The emperical formula to estimate the dergree of spin polarization at $E_F$[28] is given by Eq. (1)

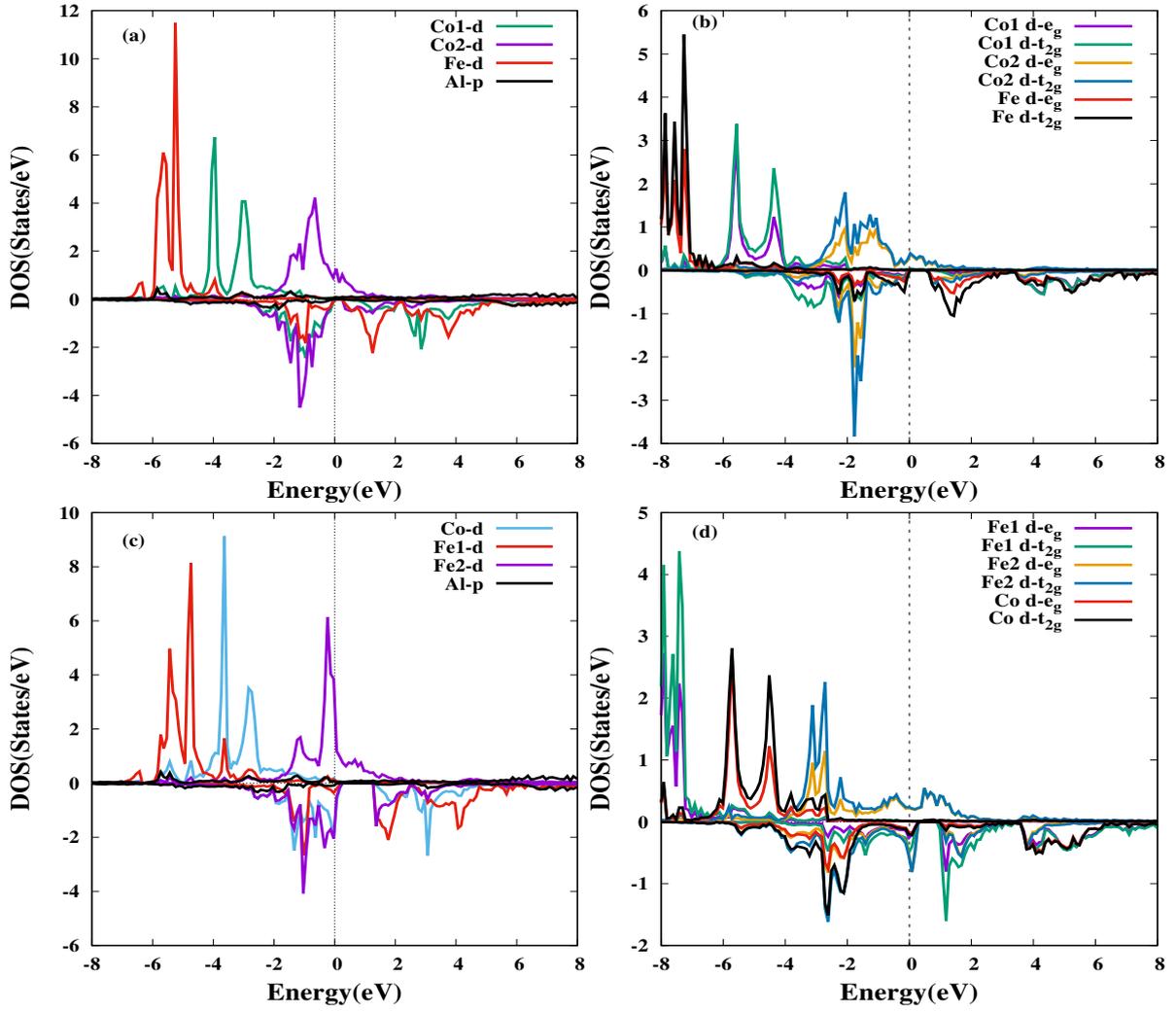

FIG. 2. Calculated partial DoS of Co$_2$FeAl: (a) GGA, (b) GGA+U and Fe$_2$CoAl:(c) GGA, (d) GGA+U

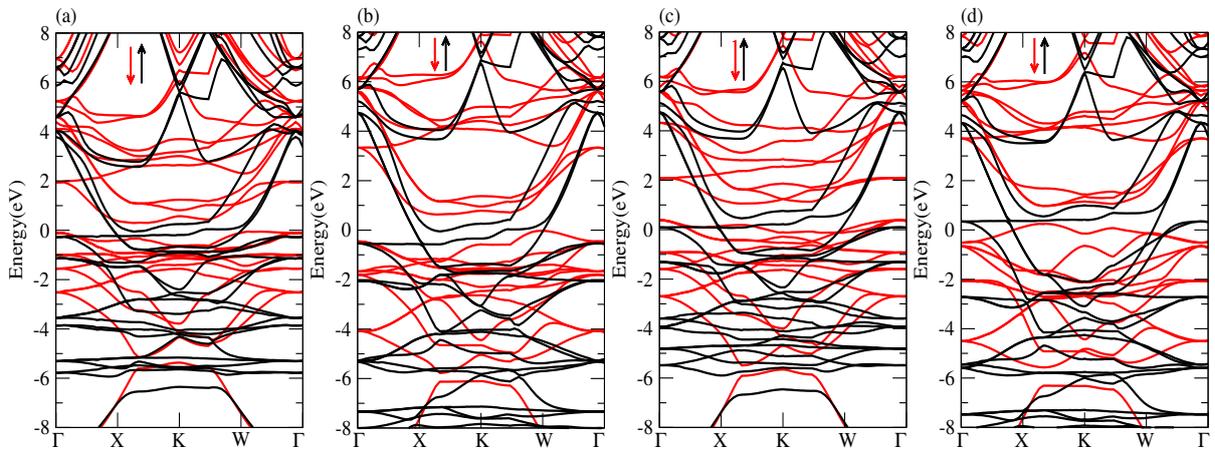

FIG. 3. Calculated band structures of Co$_2$FeAl: (a) GGA, (b) GGA+U and Fe$_2$CoAl:(a) GGA, (b) GGA+U

$$P = \frac{N_\uparrow(E_F) - N_\downarrow(E_F)}{N_\uparrow(E_F) + N_\downarrow(E_F)} \quad (1)$$

where $N_\uparrow(E_F)$ and $N_\downarrow(E_F)$ are the number of density of states at $E_F$ for spin-up and spin-down channels respectively. The partial DoS of $Fe_2CoAl$ is shown in Fig.2(c,d) which predict its metallic characteristic. The band gap in the spin-down channel is located above the Fermi level. As one can see the $E_F$ is passing through the dispersed DoS in both the spin channels, also cross checked from Fig.3(c,d). The total magnetic moment of $L2_1$ and $C1_b$ type Heusler compound is based on the Slater-Pauling (SP) rule[56]. Full-Heusler compound follows the rule of 24; $M_t = Z_t - 24$ ($\mu_B$), where $M_t$ and $Z_t$ are the total magnetic moment and total valence electrons, respectively. The integer value of $M_t$ is another important factor to determine the half-metallic behaviour. The respective values of $Z_t$ for $Co_2FeAl$ and for $Fe_2CoAl$ are 29 and 28. So the expected values of $M_t$ for $Co_2FeAl$ and $Fe_2CoAl$ are 5 ($\mu_B$) and 4 ($\mu_B$), respectively. The calculated $M_t$= 5.00 $\mu_B$ for $Co_2FeAl$, exactly an integer value that falls within the SP-rule but $M_t$= 5.12 $\mu_B$ for $Fe_2CoAl$ deviates abruptly. This is because there are 11.5 electrons instead of 12 in the spin down band and in the majority band there 16.5 electrons instead of 16. This leads to $M_t$=16.5-11.5=5.00 $\mu_B$ instead of 4.00$\mu_B$ in $Fe_2CoAl$. To restore half metallicity the variation of lattice constant may works well. This will shift the $E_F$ into the band gap and the perfect half-metal with $M_t$=4.00 $\mu_B$ may be possible. The calculated $M_t$ values of $Co_2FeAl$ and $Fe_2CoAl$ are presented in Table II along with the previous reports. Our results are in good agreement with the available data. Another interesting feature is high value of Curie temperature ($T_C$) exhibited by majority of the Heusler compounds. The Curie temperatures ($T_C$) are calculated from the following Eq. (2)[57,58]

$$T_C = 23 + 181 \times M_t \quad (2)$$

$T_C$ is strongly related to the total magnetic moment ($M_t$) thus we can get a linear relation between $T_C$ and $M_t$ (mimic $y = mx + c$) which shows that higher value of $M_t$ gives high $T_C$. We have also estimated the strength of the magnetic inter-action by calculating the exchange coupling constants $J_{ij}$ between two atoms at $i$ and $j$ sites as a function of distance [see Fig.4]. According to a classical Heisenberg model the central atoms are embedded in a ferromagnetic coherent potential approximation (CPA) medium and the exchange coupling constants are calculated by mapping the difference of total energy due to minute change in the rotations of the two moments at $i$ and $j$ sites[59]. The exchange coupling $J_{ij}$ is given by

$$J_{ij} = \frac{1}{4\pi} \int^{E_F} d(E) \Im Tr_L \{\triangle_i T_\uparrow^{ij} \triangle_j T_\downarrow^{ji}\} \quad (3)$$

where $\triangle_i = t_{i\uparrow}^{-1} - t_{j\downarrow}^{-1}$, $t_{\uparrow\downarrow}^{-1}$ is the atomic $t$-matrix of the of magnetic impurities at site $i$ for the spin up/down state. $T_{\uparrow\downarrow}^{ij}$ is the scattering path operator between $ij$ sites for the spin up/down state. $Tr_L$ is the trace over the orbital variables. The Curie temperature ($T_C$) can be calculated from the mean field approximation (MFA) by using Eq.(4).

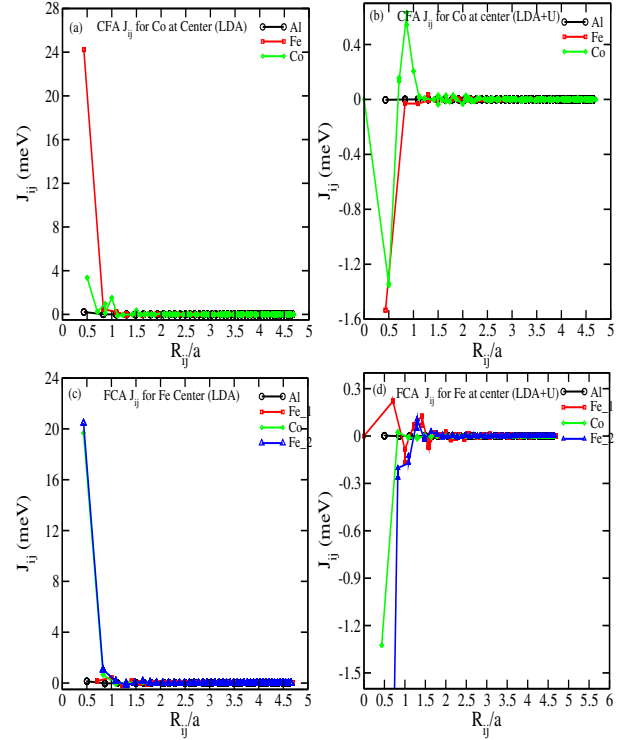

FIG. 4. Calculated exchange coupling $J_{ij}$ parameter of: (a) $Co_2FeAl$ at Co center within LDA (b) $Co_2FeAl$ at Co center within LDA+U, (c) $Fe_2CoAl$ at Fe center within LDA and (d) $Fe_2CoAl$ at Fe center within LDA+U

$$k_B T_C \langle s_i \rangle = \frac{2}{3} c \sum_{j, r \neq 0} J_{ij}^{0,r} \langle s_j \rangle \qquad (4)$$

where $c$ is the concentration of impurities and $k_B$ is the Boltzmann constant, $\langle s_j \rangle$ is the average $j$ component of the unit vector $s_r^j$ along the directon of magnetization. The calculated $T_C^{MFA}$ are tabulated in TableII and overestimated because of the inadequate description of the magnetic percolation effect[60,61].

## IV. X-RAY ABSORPTION SPECTRA (XAS) AND X-RAY MAGNETIC CIRCULAR DICHROISM (XMCD)

A theoretical study of XAS and XMCD have been performed based on the sum rule within the frame work of DFT+U in an absence of external applied field. Fig.5(a,b) shows the calculated XAS and XMCD spectra of Co and Fe at $L_{2,3}$ edges for $Co_2FeAl$ and $Fe_2CoAl$. The XAS of Co atoms in $Co_2FeAl$ are more sharp with high intensity as compared to Fe atoms. On the other hand the XAS of Fe atoms in $Fe_2CoAl$ shows more intense peaks. The $L_3$ and $L_2$ edges are denoted as peak 1 and peak 2 in Fig.5(a,b). The ratio of intensities of XAS, between $L_3$ and $L_2$ edges in Co is 2:1 and Fe is 2.4:1 for $Co_2FeAl$, in good agreement with the experiment[62]. Moreover, the ratio of $L_3$ and $L_2$ in $Fe_2CoAl$ is 3:1 for Co and 2.6:1 for Fe atom. The similar result of deviation of branching ratio from standard 2:1 as in later case has also been reported in all one-electron approach[63,64]. This discrepancy may be because of the implementation of inaccurate Coulomb potential ($U_{Co}/U_{Fe}$) for electron core-hole exchange correlation in $Fe_2CoAl$. The $L_{3,2}$-XAS of Co and Fe are related to the unoccupied d-states near $E_F$. The $L_3$ edges in Fig.5(a, b) are derived from the spin-resolved unoccupied Co-d $t_{2g}(\uparrow)$ and Fe-d $t_{2g}(\uparrow)$ states for $Co_2FeAl$ and $Fe_2CoAl$, respectively [see Fig.2 (b,d)]. These results are in good agreement with the experimental one calculated from the ultra-high vacuum magnetron sputtering by applying pressure below $8 \times 8^{-10}$ Pa[65]. The absorption peaks at 2 eV and 18 eV are due to the transition of $2p_{3/2} \rightarrow 3d$ and $2p_{1/2} \rightarrow 3d$ which corresponds to $L_3$ and $L_2$ edges, respectively. The XMCD spectra are calculated corresponds to XAS. The first XMCD spectra of Co and Fe atoms occur at 0.5 eV but in the second XMCD spectra Co edgeout Fe by $\sim$3.0 eV [Fig.5(a)]. The occurence of XMCD Co-$L_{2,3}$ and Fe-$L_{2,3}$ peaks are in a direct relation to ferromagnetic ordering of $Co_2FeAl$, which has also been reported earlier[62,65]. The similar explanation may follows for $Fe_2CoAl$ as well [Fig.5(b)].

## V. PHONON PROPERTIES

We have also presented the GGA calculation of lattice dynamics for $Co_2FeAl$ and $Fe_2CoAl$ with the relaxed structure. The graphical representation of the phonon dispersion curve within first Brillouin zone and the phonon density of states

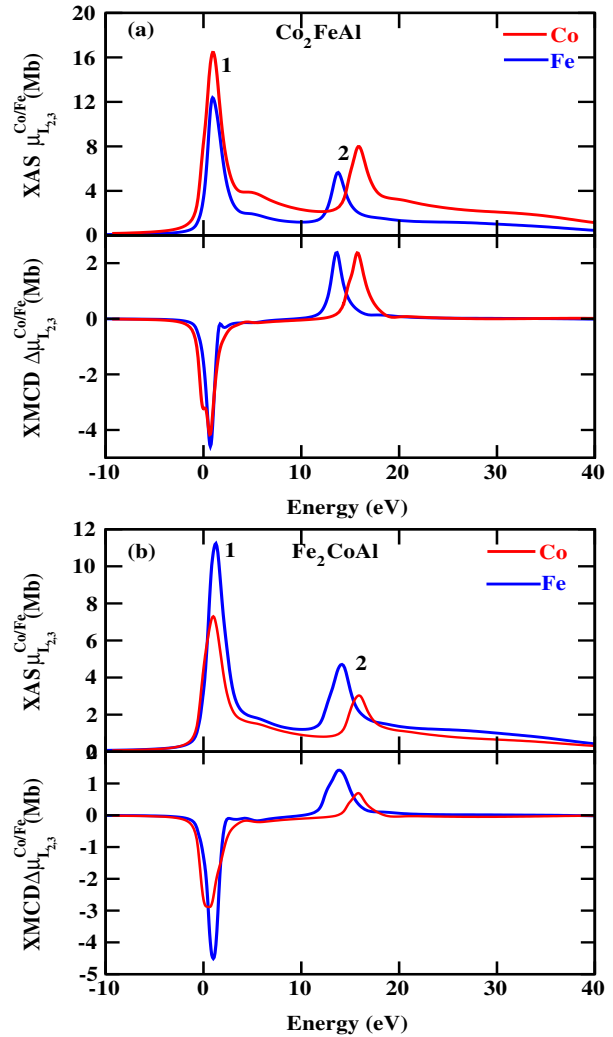

FIG. 5. XAS and XMCD of (a)$Co_2FeAl$ and (b) $Fe_2CoAl$ calculated within LDA+U

are shown in Fig.6 (a,b). Due to the presence of four atoms in a primitive cell, we can have twelve vibrational modes of phonon at any $q$-point, it consists of three acoustic modes and nine optical modes. The detail information about the dynamical stability of $Co_2FeAl$ and $Fe_2CoAl$ structure has been confirmed due to the absence of imaginary phonon frequencies[see Fig.6(a, b)]. For $Co_2FeAl$, the optical branches are located at 120.258 $cm^{-1}$, 215.194 $cm^{-1}$ and 258.68 $cm^{-1}$ at the $\Gamma$-point. For $Fe_2CoAl$, the optical branches are observed at 211.134 $cm^{-1}$, 228.466 $cm^{-1}$ and 286.239 $cm^{-1}$ along the $\Gamma - point$ [Fig.6(b)]. However, we do not have sufficient data to compare our results for the same compounds, therefore we are compelled to compare with that of the analogous compounds $Ru_2FeX$ (X=Si, Ge)[70]. The reported value of optical branches[70] at $\Gamma - point$ for $Ru_2FeSi$ are 248.319 $cm^{-1}$, 253.221 $cm^{-1}$, 273.829 $cm^{-1}$, 289.636 $cm^{-1}$ and 365.966 $cm^{-1}$ and for $Ru_2FeGe$ are 205.147 $cm^{-1}$, 215.651 $cm^{-1}$, 221.954 $cm^{-1}$, 232.458 $cm^{-1}$ and 300.210 $cm^{-1}$. In

| Compd. | $M_{Co}$ | $M_{Fe}$ | $M_t^{Cal}$ | $M_{Co}*$ | $M_{Fe}*$ | $M_t^{Rep}$ | $M_{SP}$ | $T_C^{cal}$ | $T_C^{MFA}$ | $T_C^{Expt}$ |
|---|---|---|---|---|---|---|---|---|---|---|
| CFA-GGA | 1.236 | 2.757 | 4.99 | $1.2^a$ | $2.8^a$ | $5.08^a$ | 5.0 | 926.2 | 1332.2 | $900^{d,e}$ |
|  |  |  |  |  |  | $4.82$-$5.22^b$ |  |  |  | $1000^f$ |
|  |  |  |  | $0.79^g$ | $2.77^g$ | $4.25^g$ |  |  |  | $1100^g$ |
| CFA-GGA+U | 1.290 | 2.952 | 5.00 |  |  |  | 5 | 928 |  |  |
| FCA-GGA | 1.118 | 2.559 | 5.05 | $1.0^a$ | $2.5^a$ | $5.14^a$ | 4.0 | 937.05 | 1025.4 | 700-800$^h$ |
|  | 1.64 |  |  |  | $1.6^a$ | $4.91^c$ |  |  |  |  |
| FCA-GGA+U | 0.888 | 2.759 | 5.33 |  |  |  | 4.0 | 987.73 |  |  |
|  | 2.159 |  |  |  |  |  |  |  |  |  |

TABLE II. Calculated partial magnetic moment of Co ($M_{Co}$), Fe ($M_{Fe}$) and Total Magnetic moment ($M_t$) in $\mu_B$ in comparision with the reported data ($M^{Rep}$) and Slater-Pauling rule ($M_{SP}$), Curie temperatures calculated from Eq.2 ($T_C^{cal}$) and Mean field approximation ($T_C^{MFA}$) in K, *denote other work ($a^{39}$, $b^{51}$, $c^{52}$, $d^{67}$, $e^{68}$, $f^{69}$, $g^{65}$, $h^{66}$)

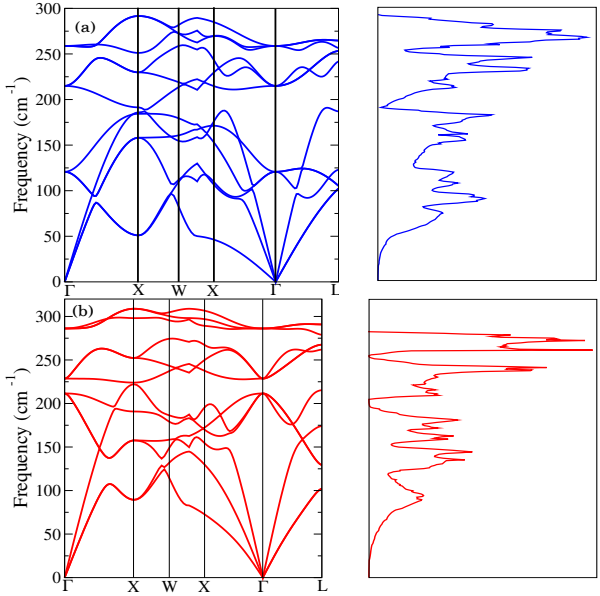

FIG. 6. Phonon dispersion & Phonon DOS of (a)Co$_2$FeAl (b) Fe$_2$CoAl

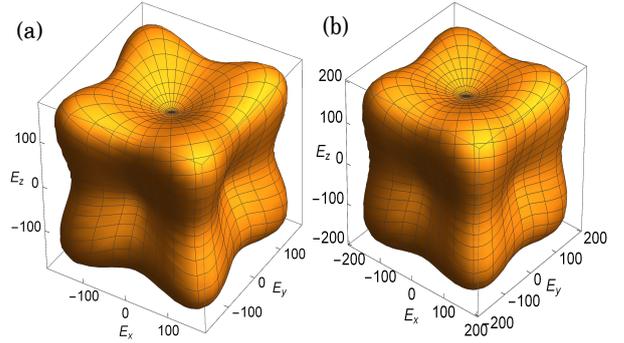

FIG. 7. The calculated 3D surface construction of Youngs modulus in (a) Co$_2$FeAl and (b) Fe$_2$CoAl

our calculation we have obtained only three optical branches at the $\Gamma-point$ and degenerated into six modes along X-point. On the other hand, there are 5 distinct optical branches at $\Gamma-point$ for Ru$_2$FeX which may be due to the presence of heavier Ru atom. As a result of heavier masses in Ru$_2$FeX, the optical and acoustic phonon couplings are strong and dominant[70]. In our case the optical and acoustic phonon coupling is less derived due to the presence of low atomic masses.

## VI. ELASTIC PROPERTIES

The validation of experimental elastic constant from the standard hypothesis gives further understanding about the possibility of synthesizes the closely resembling compounds. The computation of elastic constants and moduli of elasticity of a cubic Heusler compounds are simple and rudimentary which have been reported earlier in some of the articles[74–77]. The calculated elastic constants from cubic-elastic[50] code developed by Jamal et al. are presented in Table III. A necessary criteria for mechanical stability of the cubic crystal in relation to elastic constants are as follows;

$$C_{11} - C_{12} > 0, C_{44} > 0, (C_{11} + 2C_{12} > 0 \qquad (5)$$

Our computed data of elastic constants strictly follows the above mentioned stability conditions(Eq.5). Therefore, both Co$_2$FeAl and Fe$_2$CoAl are Mechanically stable. The value of B/G=3.32 > 2.73 suggested that Co$_2$FeAl is more ductile as compared to Fe$_2$CoAl. Poisson's ratio ($\sigma$) is an important parameter to describe the nature of atomic bonding in the crystal. The $\sigma \sim 0.1$ referes to covalent bonding, moreover our calculated $\sigma$'s are 0.36 and 0.47 predict metallic bonding. The melting temperature of Fe$_2$CoAl is 1656.92±300 K higher than that of Co$_2$FeAl, 1390±300K. The hardness of a crystalline material is also a crucial factor in determining their practical applications and can be estimated from Young's modulus (Y). Young's modulus (Y) can also provide an infromation about the stiffness of a material, larger the Y values harder to deform. The results of direction dependences of Young modulus shows minima along the [100] axis (i.e., Y$_X$), [010] axis (i.e., Y$_Y$), [001] axis (i.e., Y$_Z$) and the maxima is located along [111] [see Fig.7(a,b)].

| Compounds | $C_{11}$ | $C_{12}$ | $C_{44}$ | B | G | B/G | Y | $\sigma$ |
|---|---|---|---|---|---|---|---|---|
| *$Co_2FeAl$ | 243.09 | 141.64 | 138.59 | 209.27 | 62.86 | 3.32 | 171.426 | 0.36 |
| *$Fe_2CoAl$ | 219.60 | 186.78 | 139.17 | 208.67 | 76.39 | 2.73 | 205.546 | 0.47 |
| $Fe_2CrAl^a$ | 229.56 | 151.41 | 275.94 | 177.46 | 181.267 | 0.979 | | |
| CASTEP (GGA)[b] | 289.75 | 156.75 | 153.07 | 201.08 | 106.617 | 1.886 | | |
| CASTEP (LDA)[b] | 258.24 | 134.78 | 139.14 | 177.09 | 99.21 | 1.785 | | |
| $Co_2CrAl^a$ | 355.82 | 314.69 | 417.86 | 328.40 | 258.99 | 1.268 | | |
| $Fe_2ScAl^a$ | 280.14 | 51.03 | 120.03 | 127.40 | 117.854 | 1.081 | | |
| $Co_2ScAl^a$ | 180.16 | 162.26 | 150.86 | 168.23 | 94.141 | 1.787 | | |
| $Co_2FeAl^c$ | 250.25 | 171.38 | 148.10 | 197.67 | 87.50 | 2.259 | 113.17 | |

TABLE III. Elastic constant ($C_{ij}$), Bulk Modulus, Shear Modulus (G), Youngs' modulus (Y) are in (GPa), B/G and Poisson's ratio ($\sigma$) in comparision with the available reported data. Here, * denote our results, $a^{71}, b^{72}, c^{73}$

## VII. CONCLUSION

The electronic, magnetic, elastic and X-ray spectroscopic properties of $Co_2FeAl$ and $Fe_2CoAl$ have been computed from GGA and GGA+U functionals. Both $L2_1$ and $C1_b$ structures are stabilized by calculating the minimum energy corresponds to optimized lattice constants. The phonon dispersion relation as a function of frequency have been calculated to confirm their thermodynamical stability. The absence of imaginary phonon in the whole Brillouin zone of both $L2_1$ and $C1_b$ structures is a concrete proof of dynamical stability. Both $Co_2FeAl$ and $Fe_2CoAl$ possess strongly correlated d-electrons which can be treated more efficiently by the implementation of onsite Hubbard Coulomb potential (U) as GGA+U functional. On careful investigation of electronic properties we have found the blende of conducting and semiconducting states, a finger print of typical half-metallicity in $Co_2FeAl$. Whereas $Fe_2CoAl$ exhibit pure metallic behaviour with disperse Co-d and Fe d-states around the $E_F$. An integer value of total magnetic moment (5.0 $\mu_B$) is an additional testament of the half-metallicity in $Co_2FeAl$ in accordance with the Slater-Pauling rule. However, for $Fe_2CoAl$ the total magnetic moment deviate from the integer value defying the Slater-Pauling rule. We have found an enhanced band gap in $Co_2FeAl$ with the application of GGA+U. Wide band gap half-metals with high value of $T_C$($\sim$1000 K) are critical factor for spintronic technology. The calculated values of $T_C$'s from Eq.2 are in close agreement with the experimental data where as the results from mean field approximation (MFA) are overestimated. The XAS and XMCD spectra are also calculated from LDA+U. Our result of XAS and XMCD spectra of $Co_2FeAl$ are consistent with the avaiable data. The $L_3$:$L_2$ ratio deviates from 2:1 branching ratio in $Fe_2CoAl$ may be due to inaccurate choice of Coulomb potential (U). The elastic properties are highly crucial for the industrial application. The high value of $B/G > 2.0$, high $\sigma$ and high melting temperature are credentials for practical application in device fabrication at high temperature.

## VIII. ACKNOWLWDGEMENT

This project is jointly supported by DST & RFBR (under Indo-Russian joint project). Dr. D. P. Rai acknowledges Department of Science and Technology (DST) New Delhi, Govt. of India vide Lett. No. INT/RUS/RFBR/P-264. Prof. Gennady M. Mikhailov acknowledges Russian Foundation for Basic Research, RFBR 17-57-45024.